\documentclass[twoside,leqno,twocolumn]{article}

\usepackage[letterpaper]{geometry}
\usepackage{ltexpprt}
\usepackage[utf8]{inputenc}
\usepackage{hyperref}
\usepackage{url}
\usepackage{booktabs}
\usepackage{multirow}
\usepackage{amsfonts}
\usepackage{amsmath}
\usepackage{bm}
\usepackage{nicefrac}
\usepackage{microtype}
\usepackage[table]{xcolor}
\usepackage{gnuplottex}
\usepackage{printlen}
\usepackage{enumerate}
\usepackage{algorithm}
\usepackage{algpseudocode}
\usepackage{balance}
\newcommand{\sigcell}[1]{\cellcolor{gray!25}{#1}}

\newcommand{\iid}{\mathbin{\overset{\text{\tiny{iid}}}{\sim}}}
\DeclareMathOperator{\expectation}{\operatorname{\mathbb{E}}}
\DeclareMathOperator{\variance}{\operatorname{\mathbb{V}}}

\DeclareMathOperator{\entropy}{\operatorname{\mathbb{H}}}

\DeclareMathOperator{\causent}{\operatorname{\mathbb{C}}}

\newtheorem{definition}{Definition}[section]
\newtheorem{problem}{Problem}[section]

\begin{document}

\title{Two-Sample Testing for Event Impacts in Time Series}

\author{Erik Scharw\"achter\thanks{
Chair of Data Science and Data Engineering,
Bonn-Aachen International Center for Information Technology,
University of Bonn, Germany,
\{scharwaechter,mueller\}@bit.uni-bonn.de} \\
\and
Emmanuel M\"uller}

\date{}

\maketitle
\fancyfoot[R]{\scriptsize{Copyright \textcopyright\ 2020 by SIAM\\
Unauthorized reproduction of this article is prohibited}}

\begin{abstract} \small\baselineskip=9pt
In many application domains, time series are monitored to detect extreme events
like technical faults, natural disasters, or disease outbreaks.
Unfortunately, it is often non-trivial to select both a time series that is informative about
events and a powerful detection algorithm:
detection may fail because the detection algorithm is not suitable, or because
there is no shared information between the time series and the events of interest.
In this work, we thus propose a non-parametric statistical test for shared information
between a time series and a series of observed events.
Our test allows identifying time series that carry information on event occurrences
without committing to a specific event detection methodology.
In a nutshell, we test for divergences of the value distributions of the time series
at increasing lags after event occurrences with a multiple two-sample testing approach.
In contrast to related tests, our approach is applicable for time series over arbitrary
domains, including multivariate numeric, strings or graphs.
We perform a large-scale simulation study
to show that it outperforms or is on par with related tests on our task for univariate
time series.
We also demonstrate the real-world applicability of our approach on datasets
from social media and smart home environments.

\end{abstract}

\section{Introduction}

Event detection in time series is an active research topic for at least two decades
\cite{Guralnik1999,Ihler2006,Sakaki2010,Amodeo2011,Lorey2011,Tsytsarau2014a}.
Typical event detection algorithms monitor a time series for anomalies, extreme values
or changes in the probability distribution, in the hope that these patterns coincide with
some exogenous event of interest.
Prominent examples are the detection of earthquakes
\cite{Sakaki2010,Earle2011,Robinson2013}
and public health issues \cite{Paul2011,Kanhabua2012b}
from social media time series.
The fundamental assumption of any event detection method is that there is a statistical
association between the behavior of the time series and the occurrence of events:
if the time series and the event series are statistically independent, it is impossible
to detect events by observing the time series.

\begin{figure*}[tb]
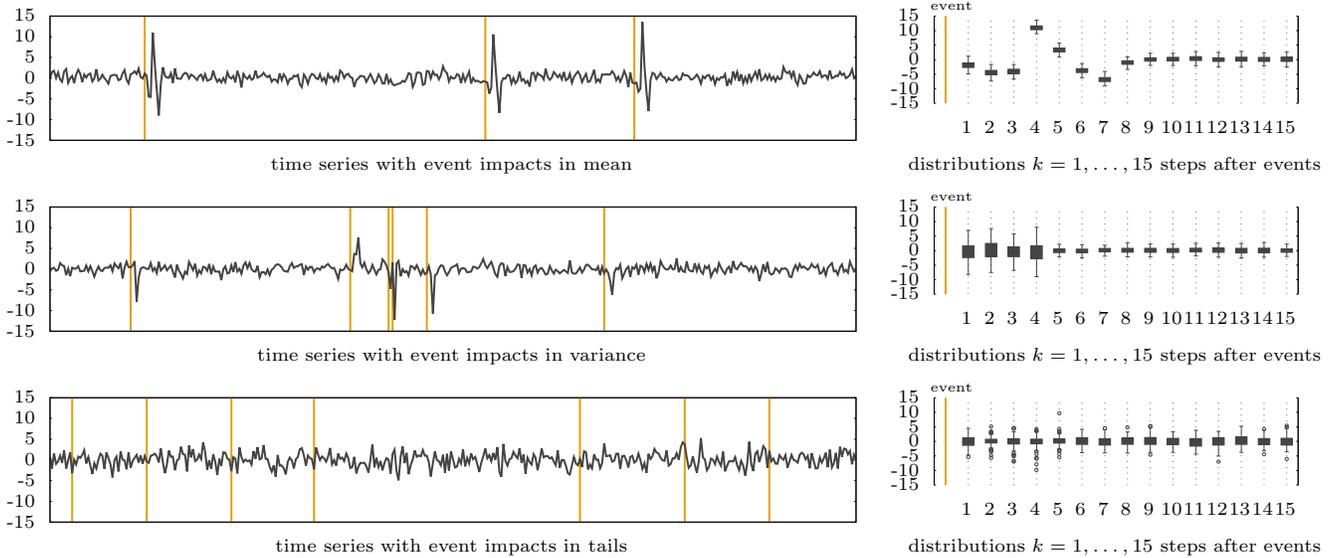

\begin{center}
\begin{gnuplot}[terminal=cairolatex,terminaloptions={size 6.80914,3.0}]
set multiplot layout 3,2
unset xlabel
unset ylabel

unset xtics
set key off
set format x '\scriptsize{set format y '\scriptsize{set ytics scale 0.2
set yrange [-15:15]
set bars 0.2

set lmargin at screen 0.02; set rmargin at screen 0.64
set xrange [100:500]
set xlabel '\scriptsize{time series with event impacts in mean}'
plot "data/ex-filter-ests.dat" u 1:($2*16) w i lc 4 lw 3 notitle, \
     "data/ex-filter-ests.dat" u 1:($2*-16) w i lc 4 lw 3 notitle, \
     "data/ex-filter-ests.dat" u 1:3 w l lc rgb '#444444' lw 3
set lmargin at screen 0.70; set rmargin at screen 0.98
set xrange [-0.5:15.5]
set xlabel '\scriptsize{distributions $k=1,\dotsc,15$ steps after events}'
set xtics 1,1 nomirror scale 0
set grid xtics mxtics
set style boxplot nooutliers labels off
set style data boxplot
set style fill solid noborder
set border 10
set arrow from 0,-15 to 0,15 lw 3 lc 4 nohead
set label '\tiny{event}' at -0.6,19 textcolor ls 4
plot "data/ex-filter-ests-trunc.dat" u (1):3:(0.5):4 notitle lc rgb '#444444'

unset grid
unset xtics
unset mxtics
set border 15
set lmargin at screen 0.02; set rmargin at screen 0.64
set xrange [100:500]
set xlabel '\scriptsize{time series with event impacts in variance}'
plot "data/ex-var-ests.dat" u 1:($2*16) w i lc 4 lw 3 notitle, \
     "data/ex-var-ests.dat" u 1:($2*-16) w i lc 4 lw 3 notitle, \
     "data/ex-var-ests.dat" u 1:3 w l lc rgb '#444444' lw 3
unset xrange
set lmargin at screen 0.70; set rmargin at screen 0.98
set xrange [-0.5:15.5]
set xlabel '\scriptsize{distributions $k=1,\dotsc,15$ steps after events}'
set xtics 1,1 nomirror scale 0
set grid xtics mxtics
set style boxplot nooutliers labels off
set style data boxplot
set style fill solid noborder
set border 10
plot "data/ex-var-ests-trunc.dat" u (1):3:(0.5):4 notitle lc rgb '#444444'

unset grid
unset xtics
unset mxtics
set border 15
set lmargin at screen 0.02; set rmargin at screen 0.64
set xrange [500:900]
set xlabel '\scriptsize{time series with event impacts in tails}'
plot "data/ex-tail-ests.dat" u 1:($2*16) w i lc 4 lw 3 notitle, \
     "data/ex-tail-ests.dat" u 1:($2*-16) w i lc 4 lw 3 notitle, \
     "data/ex-tail-ests.dat" u 1:3 w l lc rgb '#444444' lw 3
set style rect fc rgb '#ffffff' fs solid 1.0 noborder
set obj rect from screen 0.69, graph 0 to screen 0.695, graph 1 front
set obj rect from screen 0.985, graph 0 to screen 1, graph 1 front
unset xrange
set lmargin at screen 0.70; set rmargin at screen 0.98
set xrange [-0.5:15.5]
set xlabel '\scriptsize{distributions $k=1,\dotsc,15$ steps after events}'
set xtics 1,1 nomirror scale 0
set grid xtics mxtics
set style boxplot outliers pt 6 labels off
set style data boxplot
set style fill solid noborder
set pointsize 0.2
set border 10
plot "data/ex-tail-ests.dat" u (1):3:(0.5):4 notitle lc rgb '#444444'
\end{gnuplot}
\caption{Different types of event impacts in a time series. Vertical lines (orange) indicate
event occurrences. The boxplots on the right depict, for every time series, the empirical conditional
value distributions, given that the last event occurred $k$ time steps ago:
$\mathbb{P}(X_t \mid E_{t-k}=1,E_{t-k+1}=0,...,E_{t}=0)$.}
\label{figEventConditionalScatter}
\end{center}
\end{figure*}

In practice, there are numerous ways in which a time series and an event series can be
statistically associated. Some associations are easy to exploit for event detection,
others require more advanced technologies or cannot be exploited effectively.
Figure~\ref{figEventConditionalScatter} shows three example pairs of event series and
time series, where each pair is coupled differently.
In the simplest case, events lead to temporary fluctuations of the mean of the time series,
as illustrated in Figure~\ref{figEventConditionalScatter}~(top left).
Every event occurrence induces the same shape in the time series.
The boxplots in Figure~\ref{figEventConditionalScatter}~(top right) summarize the value
distributions of the time series given that the last event occurrence was $k=1,...,15$ time
steps ago.
They show that the mean varies for a few time steps and then stabilizes.
However, events can have more subtle effects.
In Figure~\ref{figEventConditionalScatter}~(middle row), events temporarily increase
the variance of the time series---as indicated by wider boxes and whiskers in the first few boxplots.
In Figure~\ref{figEventConditionalScatter}~(bottom row), events increase the risk of
extreme observations from the tails of the distribution---as indicated by a larger number
of outliers in the first few boxplots.
Such visual analyses are limited to univariate numeric time series.
If we consider multivariate numeric time series, or time series of strings or graphs,
it is unclear how to proceed visually, and quantitative statistical methods are required.

A natural way to assess whether there is a statistical relation between past event
occurrences and present values of the time series is to perform a statistical test
for causality, e.g., Granger causality \cite{Granger1969} or non-zero
transfer entropy \cite{Schreiber2000}.
However, existing tests are restricted to univariate time series, to impacts in mean,
or have estimation issues.
We thus propose a novel \textbf{statistical independence test} between
the current value of the time series and past values of the event series.
Our test can be embedded in the information-theoretic framework of causation entropy \cite{Sun2014a}
that generalizes Granger causality and transfer entropy.
Algorithmically, we test for independence by testing for pairwise divergences in the
distributions of the time series at increasing lags after event occurrences.
This allows us to leverage recent advancements in \textbf{two-sample testing} \cite{Gretton2012},
and makes our test applicable to time series from arbitrary domains, including multivariate numeric,
string or graph data.
In a \textbf{large scale simulation study}, we evaluate the power of our test against tests for
Granger mean causality and non-zero transfer entropy.
Furthermore, we demonstrate the real-world applicability of our test with use cases from social media
analysis and household electricity monitoring.

\section{Related work}
\label{secRelatedWork}

\paragraph{Causal inference.}
Our closest competitors are tests for causal inference in time series.
Granger causality \cite{Granger1969} and transfer entropy \cite{Schreiber2000}
are notions of statistical association between time series used to identify cause-effect
relationships. Same as our test, they can be subsumed in the framework of causation
entropy \cite{Sun2014a}. We include them as competitors in our experimental evaluation.
Both assume that the target time series is \textbf{univariate}, and can only be tested independently
per dimension on multivariate target time series.
Traditionally, Granger causality tests focus on the conditional mean of the time series
and utilize a likelihood ratio statistic based on vector autoregressive models.
More efficient estimators have been developed, e.g., based on state-space models \cite{Barnett2015},
based on kernel regressions to capture non-linear couplings \cite{Marinazzo2008},
and other nonparametric predictors \cite{Bell1996}.
By design, they all \textbf{fail to capture causal effects that
do not alter the conditional mean of the distribution}.
Departing from causality in mean, a few nonparametric tests for \emph{general-sense}
Granger causality,---not restricted to the conditional mean---, have
been proposed \cite{Hiemstra1994,Diks2006,Nishiyama2011},
However, the vast majority of tests are established for \textbf{real-valued time series only}:
it is unclear how they perform on time series over other domains.
Our approach is not restricted to real-valued time series, but \textbf{operates on all
types of data}, if a two-sample test is available.
A notable exception are tests based on transfer entropy \cite{Bossomaier2016}:
they directly operate on the conditional distributions and are thus nonparametric \emph{and}
applicable for numeric and categorical data.
Transfer entropy measures information flow between time series, and can be used as a nonparametric
statistic to test for general-sense Granger causality.
However, transfer entropy inherits the \textbf{difficulties in estimation} of mutual information
and entropy \cite{Bossomaier2016}, which limits its detection performance.
Our approach is nonparametric \emph{and} has a high detection performance.

\paragraph{Two-sample testing.}
Methodologically, our test heavily relies on multiple two-sample testing.
In two-sample testing, the problem is to decide whether two random samples come from
the same probability distribution, or from different distributions.
The most well-known two-sample test is Student's t-test that compares the means of two
distributions. For univariate continuous data, the Kolmogorov-Smirnov (KS) two-sample test
compares the complete empirical distribution functions \cite{Wasserman2004}, but suffers
from estimation issues.
Recently, kernel-based approaches to two-sample testing have been developed \cite{Gretton2012,Gretton2009,Gretton2012a,Zaremba2013}
that are applicable for arbitrary domains.
A more comprehensive review of two-sample testing can be found in \cite{Gretton2012}.

\section{Independence Problem}
\label{secProblem}

\subsection{Terminology.}
A \textit{time series} is a random process $\mathcal{X} = \{X_t\}_{t \in \mathbb{Z}}$ where
all $X_t$ for $t \in \mathbb{Z}$ are random variables.
An \textit{event series} $\mathcal{E} = \{E_t\}_{t\in\mathbb{Z}}$ is a specific time series with
discrete random variables $E_t$ that take only the values 0 and 1.
The outcome $E_t=1$ indicates that there is an event at time $t$.
A random process $\mathcal{Z}$ is \textit{stationary} if for all $k \in \mathbb{N}$ and
$t_1,...,t_k \in \mathbb{Z}$, the joint probability density function (or joint probability mass
function in case of discrete outcomes) of $Z_{t_1},...,Z_{t_k}$ is shift invariant,
i.e., $\mathbb{P}(Z_{t_1},...,Z_{t_k}) = \mathbb{P}(Z_{t_1+h},...,Z_{t_k+h})$ for all
$h \in \mathbb{Z}$.
Two processes $\mathcal{X}$ and $\mathcal{E}$ are \textit{jointly stationary}
if the bivariate process $\{(X_t, E_t)\}_{t \in \mathbb{Z}}$ is stationary.
Throughout this work, we make the standard assumption that $\mathcal{X}$ and $\mathcal{E}$ are
jointly stationary, such that their statistical association can be estimated from a
single observed pair.

\subsection{Problem statement.}
Let $\mathcal{X}_{{}<t}$ and $\mathcal{E}_{{}<t}$ denote the histories of the two
series up to time $t-1$.
The histories may be cropped at some lag $l$, such that
$\mathcal{X}_{{}<t} = \mathcal{X}_{{}<t}^{(l)} := \{X_{t-l},...,X_{t-1}\}$,
and analogously for $\mathcal{E}_{{}<t}$.
We address the following hypothesis testing problem:
\begin{problem}
\label{problemEIdetection}
Given a time series $\mathcal{X}$ and an event series $\mathcal{E}$,
test the null hypothesis
\begin{equation}
\label{eqnCondInd}
H_0: \mathbb{P}(X_t \mid \mathcal{E}_{{}< t}) = \mathbb{P}(X_t)
\end{equation}
against the alternative hypothesis
\begin{equation}
\label{eqnNoCondInd}
H_1: \mathbb{P}(X_t \mid \mathcal{E}_{{}< t}) \neq \mathbb{P}(X_t).
\end{equation}
\end{problem}

Problem~\ref{problemEIdetection} is an independence test between a single random variable
$X_t$ and a set of random variables $\mathcal{E}_{{}<t} = \{E_{t-l}, ..., E_{t-1}\}$.
The challenge is to efficiently test this independence without making restricting
assumptions on the domain of $\mathcal{X}$, and avoiding estimation issues that limit
the practical applicability.

\subsection{A family of tests.}
The test above belongs to a family of tests subsumed under the information-theoretic
framework of causation entropy \cite{Sun2014a}.
Let $\mathcal{X}$ and $\mathcal{Y}$ be two time series, and $\bm{\mathcal{S}}$ be a set
of time series.
Causation entropy is a measure for information flow from $\mathcal{Y}$ to $\mathcal{X}$,
taking all additional information from the set $\bm{\mathcal{S}}$ into account.
If there is no information flow, the time series are conditionally independent.
Formally, let $\mathbb{H}(\cdot \mid \cdot)$ denote the conditional entropy \cite{Cover2006e}.
\begin{definition}[Causation entropy]
\label{defCausEnt}
The causation entropy from $\mathcal{Y}$ to $\mathcal{X}$ conditioned on the set of time series
$\bm{\mathcal{S}}$ is the conditional mutual information of $X_t$ and $\mathcal{Y}_{{}<t}$
given $\bm{\mathcal{S}}_{{}<t}$:
\begin{align}
\label{eqnCausEnt}
\causent_{\mathcal{Y}\rightarrow\mathcal{X}\mid\bm{\mathcal{S}}}
&:= \entropy[X_t \mid \bm{\mathcal{S}}_{{}<t}]
       - \entropy[X_t \mid \bm{\mathcal{S}}_{{}<t},\mathcal{Y}_{{}<t}].
\end{align}
\end{definition}
The causation entropy $\causent_{\mathcal{Y}\rightarrow\mathcal{X}\mid\bm{\mathcal{S}}}$ is zero
if and only if the conditional independence
\begin{equation}
\mathbb{P}(X_t \mid \mathcal{Y}_{{}<t}, \bm{\mathcal{S}}_{{}<t})
= \mathbb{P}(X_t \mid \bm{\mathcal{S}}_{{}<t})
\end{equation}
holds. If the conditional independence does not hold, the causation entropy is positive.

Different choices of the set of conditioning time series $\bm{\mathcal{S}}$ result in different
independence tests. In transfer entropy and Granger causality, the target time series itself
is used in the condition, i.e., $\bm{\mathcal{S}} = \{\mathcal{X}\}$.
Additional time series may be included in $\bm{\mathcal{S}}$ to take potential confounding
factors into account, but this makes estimation harder.
With $\bm{\mathcal{S}} = \emptyset$ and $\mathcal{Y} = \mathcal{E}$, we obtain the independence
test in Problem~\ref{problemEIdetection}.
From an information theoretic point of view, we thus test for non-zero
\emph{unconditional} causation entropy from the event series to the time series.
By employing an empty set of conditions, our test explicitly ignores the effect of confounding
factors to increase sensitivity.
The detected associations may be indirect or due to common drivers---but still
useful for event detection.

\section{Two-Sample Test Approach}
\label{secEventImpactTest}

Our approach exploits the binary nature of the event series $\mathcal{E}$ to
solve Problem~\ref{problemEIdetection} heuristically.
To this end, we apply a fundamental independence criterion for mixed random variables.
We start with the general idea and provide technical details below.
Independence of mixed random variables can be characterized
by equality of all conditional probability density functions:

\begin{theorem}[\cite{Wasserman2004}]
\label{thmIndMixedVars}
Let $A$ and $B$ be random variables, where $A$ is continuous
and $B$ is discrete with $K$ outcomes $0,...,K-1$.
$A$ is independent of $B$, if and only if all conditional probability density
functions are identical:
\begin{equation}
\begin{split}
& \mathbb{P}(A \mid B) = \mathbb{P}(A)\\
\Leftrightarrow{}&\mathbb{P}(A \mid B = 0) = ... = \mathbb{P}(A \mid B = K-1).
\end{split}
\end{equation}
\end{theorem}

Independence of $A$ and $B$ may thus be assessed by pairwise comparisons of the conditional
distribution functions.
Given a sample of independent and identically distributed (i.i.d.) pairs from $A$ and $B$,
the conditional distributions can be compared with multiple pairwise two-sample tests.
If any of the two-sample tests finds significant evidence that the two underlying
conditional distributions differ, the null hypothesis of independence must be rejected.

\subsection{Naive approach.}
Mapping this idea into our problem setting, we could naively encode the event history
$\mathcal{E}_{{}<t} = \{E_{t-l},...,E_{t-1}\}$ as a single discrete random variable with $K=2^l$
possible outcomes. The original outcome $E_{t-l}=\epsilon_{t-l},...,E_{t-1}=\epsilon_{t-1}$ with
$\epsilon_\tau \in \{0,1\}$ would then correspond to the base-2 number
$(\epsilon_{t-l},...,\epsilon_{t-1})_2 \in \{0,...,2^l-1\}$ in the novel encoding.
For a fixed lag $l$, we could then directly apply Theorem~\ref{thmIndMixedVars}
to test the independence in Problem~\ref{problemEIdetection} by obtaining i.i.d. samples
from the $2^l$ conditional distribution functions and testing for pairwise equality.
However, with this naive approach, we will run into two severe estimation problems:
(1)~The exponential number of possible outcomes means that a large number of tests have to
be performed, which reduces the detection performance. (2)~Event series are usually sparse,
meaning that many outcomes will never be realized, and no i.i.d. samples can be obtained.
The naive approach is thus not operational.

\subsection{Reducing the number of tests.}
A key idea of our independence test is that we can detect an association between the past
of the event series and the current value of the time series without testing \emph{all}
conditional distributions for divergences.
Formally, let
\begin{equation}
F^K_{\epsilon_0,...,\epsilon_{K}}
:= \mathbb{P}(X_t \mid E_{t-K}=\epsilon_0,...,E_{t}=\epsilon_{K})
\end{equation}
denote the event-conditional distribution function of order $K \in \mathbb{N}$ for an outcome
$\epsilon_0,...,\epsilon_{K}$.
For a fixed $K$, there are $2^{K+1}$ such distribution functions, many of which are not realized
in practical instances with sparse events.
For increasing $k=0,1,2,...$, the specific distribution functions $F^k_{1,0,...,0}$ describe the
conditional distributions of $X_t$ given that the most recent event happened $k$ time steps ago.
These distributions are always realized in practical instances as soon as there is a single event
in $\mathcal{E}$. The number of samples per distribution $F^k_{1,0,...,0}$ directly corresponds
to the number of events in $\mathcal{E}$.
The boxplots in Figure~\ref{figEventConditionalScatter} (right) depict these distributions
for different kinds of impacts.

We assume that events have a strong association with observations that follow immediately
in the time series, and little to no association with observations that are far away.
We thus propose to test \emph{only} the event-conditional distribution functions
$F^k_{1,0,...,0}$ with $k=0,...,K$ for divergences, where $K \in \mathbb{N}$ is some upper limit.
If all of these distributions are identical, there is no evidence for a statistical
association between the event series and the time series.
If any pair of these distribution functions diverges, we reject the null hypothesis of
independence in favor of shared information.
Formally, we simplify the hypotheses from Problem~\ref{problemEIdetection} and test
\begin{equation}
H'_0 : F^0_{1} = F^1_{1,0} = ... = F^{K}_{1,0,...,0}
\end{equation}
versus $H'_1 : \neg H'_0$.
By focusing on this specific selection of conditional distributions, we address
both estimation issues mentioned above: we decrease the number of conditional
distributions to compare from $2^{K+1}$ to $K+1$, and
we work with conditional distributions that are realized for sparse event series.
Since we ignore many conditional distributions, the resulting test procedure does
not solve Problem~\ref{problemEIdetection} exactly, but heuristically.

\subsection{Multiple test procedure.}
We test the pair of hypotheses $H'_0$ and $H'_1$ from above with the multiple
test procedure specified in Algorithm~\ref{algEITEST}. We refer to our test
as EITEST (\textbf{E}vent \textbf{I}nformation \textbf{TEST}).
The input is a realized pair of time series $\mathcal{X}=\{x_1,...,x_T\}$ and
event series $\mathcal{E}=\{e_1,...,e_T\}$
with $N=\sum e_t$ events, along with the maximum lag parameter $K$.
The output of the algorithm is a p-value.
If the p-value is smaller than the desired significance level $\alpha$,
we reject $H'_0$ in favor of $H'_1$.
In line~3, samples $\mathcal{T}_k$ from the event-conditional distribution
functions $F^k_{1,0,...,0}$ are obtained. In line~7, the pairwise two-sample
tests are called, where the output of the two-sample test is a p-value.
In lines~11 and~12, the obtained p-values are corrected for multiple testing with
Simes adjustments \cite{Dmitrienko2010}.
Details on sample construction and error rate control are given below.
The complexity is $O(KT + K^2(g(N) \cdot \log K))$, where $g(N)$ is the complexity
of the underlying two-sample test. $g(\cdot)$ is a function of $N$ since all
samples $\mathcal{T}_k$ contain at most $N$ observations.
Typically, $K$ is a small constant $K \ll T$, and the event series is sparse
with $N \ll T$. The total complexity is thus asymptotically dominated by a term
that is linear in $T$, which makes EITEST highly computationally efficient
for long time series and event series.

\begin{algorithm}[t]
\begin{algorithmic}[1]
\scriptsize
\Function{EITEST}{$\mathcal{X}, \mathcal{E}, K$}
\For{$k = 0,...,K$}
  \State $\displaystyle \mathcal{T}_k
:= \left\{x_t \mid e_{t-k}=1, e_{t-k+1}=0, ..., e_t=0\right\}$
\EndFor
\For{$i = 0,...,K-1$}
  \For{$j = i+1,...,K$}
    \State $p_{ij} := \Call{TwoSampleTest}{\mathcal{T}_i,\mathcal{T}_j}$
  \EndFor
\EndFor
\State $M := (K \cdot (K+1))/2$
\State $\hat{p}_1, ..., \hat{p}_M := \Call{SortIncreasing}{\{p_{ij} \mid i<j\}}$
\State \Return $\min_m\left\{\frac{M}{m} \cdot \hat{p}_m\right\}$
\EndFunction
\end{algorithmic}
\caption{Multiple test procedure}
\label{algEITEST}
\end{algorithm}

\subsection{Sampling the distributions.}
The two-sample tests require i.i.d.\ samples from the distribution functions $F^k_{1,0,...,0}$.
Any observation $x_t$ with $e_{t-k}=1, e_{t-k+1}=0, ..., e_{t}=0$ is a realized value
from the distribution $F^k_{1,0,...,0}$.
In line~3, we thus obtain disjoint samples by assigning observations from the
time series to $K$ subsets $\mathcal{T}_k$ such that every value $x_t$ is
assigned to $\mathcal{T}_k$ if and only if $e_{t-k}=1$ and $e_{t-k+1}=0,...,e_{t}=0$.
However, the individual observations in $\mathcal{T}_k$ are not, in general, independent.
In practice, even if two random variables $X_t$ and $X_{t'}$ from $\mathcal{X}$ are not strictly
conditionally independent given $E_{t-k},...,E_{t-1}$ and $E_{t'-k},...,E_{t'-1}$,
long-range dependencies are often weak, i.e.,
\begin{equation*}
\begin{split}
&\mathbb{P}(X_t,X_{t'} \mid E_{t-k},...,E_{t-1}, E_{t'-k}, ..., E_{t'-1}) \\
\approx{}& \mathbb{P}(X_t \mid E_{t-k},...,E_{t-1})
    \cdot \mathbb{P}(X_{t'} \mid E_{t'-k},...,E_{t'-1})
\end{split}
\end{equation*}
for all $|t-t'| > l$ with some large $l>k$. In other words, $X_t$ and $X_{t'}$ are
\emph{approximately} independent if they are far enough apart.
If serial dependencies are an issue, additional constraints can be imposed to ensure
hard minimum distances between individual observations within a set $\mathcal{T}_k$
as well as between observations across pairs of sets $\mathcal{T}_k$ and $\mathcal{T}_{k'}$.

\subsection{Controlling the family-wise error rate.}
In any statistical hypothesis test, the false positive rate is controlled
at significance level $\alpha$ by rejecting the null hypothesis only if the p-value returned
by the test is smaller than $\alpha$. In standard testing problems (no multiple testing),
the p-value is directly computed from a test statistic $T$ that collects evidence against
the null hypothesis.
The p-value specifies the probability of obtaining a value of $T$ at least as extreme as
the observed one, under the assumption that the null hypothesis is true.
When performing multiple hypothesis tests, we obtain many p-values: one for every test.
We need a procedure that rejects the individual null hypotheses
such that the false positive rate of the \emph{complete} null hypothesis is controlled at
level $\alpha$---not the false positive rate of the individual tests.
In our case, we have a family of individual null hypotheses
\begin{equation}
G_0^{i,j} : F^i_{1,0,...,0} = F^{j}_{1,0,...,0}
\end{equation}
for $0 < i < j \le K$, with alternative hypotheses
\begin{equation}
G_1^{i,j} : F^i_{1,0,...,0} \neq F^{j}_{1,0,...,0}.
\end{equation}
The \emph{complete} null hypothesis $H'_0$ is that all of the null hypotheses from the family
are \emph{simultaneously} true.
If \emph{any} of the individual null hypotheses is rejected, $H'_0$ is rejected
in favor of shared information.
We do not care which of the null hypotheses is false.
In this scenario, the family-wise error rate (FWER) \cite{Dmitrienko2010} is a suitable choice
for the false positive rate of the complete null hypothesis.

Formally, let $\mathcal{G} = \left\{G_0^{i,j} \mid 0 \le i < j \le K\right\}$ be the set of
all null hypotheses,
$\mathcal{T} \subseteq \mathcal{G}$ be the set of \emph{true} null hypotheses and
$\mathcal{R} \subseteq \mathcal{G}$ be the set of null hypotheses \emph{rejected}
by some procedure.
The FWER is the probability that at least one of the true null hypotheses is rejected, i.e.,
$\mathbb{P}(\mathcal{T} \cap \mathcal{R} \neq \emptyset)$ \cite{Dudoit2007}.
To guarantee $\mathbb{P}(\mathcal{T} \cap \mathcal{R} \neq \emptyset) < \alpha$, we use Simes
adjustments \cite{Dmitrienko2010}.
Let $M := |\mathcal{G}| = K \cdot (K+1) / 2$ be the total number of pairwise two-sample tests,
and $\hat{p}_{1}, ..., \hat{p}_{M}$ be the p-values returned by the tests, ordered increasingly.
We reject the complete null hypothesis $H'_0$ if $\hat{p}_{m} < \frac{m}{M} \alpha$
for \emph{any} $m = 1,...,M$.
The corresponding adjusted $p$-value for the multiple test decision can be obtained from
the individual $p$-values as $\min_{m} \{\frac{M}{m} \hat{p}_{m}\}$.

\section{Experiments}

We evaluate EITEST against the standard Granger causality test based on
VAR models (GC-VAR) and a test for non-zero transfer entropy (TE-KSG).
We perform a large-scale simulation study, where we assess the performance of all
approaches on coupled pairs of time series and event series, generated by different
event impact models.
We also generate uncoupled pairs by randomly permuting the event series after
generating a coupled pair.
To assess the detection performance of all approaches, we report their true positive
and false positive rates.
At last, we demonstrate the utility of our test with two real-life applications.

\paragraph{Evaluation measures.}
A true positive is a coupled pair of time series and event series, generated by any of
the event impact models described below, that is correctly detected as being coupled.
A false positive is an uncoupled pair that is falsely detected as being coupled.
The corresponding true positive rate (TPR, power) and false positive rate (FPR) are obtained
by normalizing over the total number of coupled and uncoupled pairs, respectively.
TPR should ideally be close to 1, whereas the FPR should be upper bounded by
the significance level $\alpha$ that was chosen for the test.

\paragraph{Setup.}
We set the significance level to $\alpha = 0.05$.
In EITEST, we use the maximum lag $K=32$. We report results with the Kolmogorov-Smirnov (KS)
two-sample test \cite{Wasserman2004} and the Maximum Mean Discrepancy (MMD) test \cite{Gretton2012}
with default RBF kernel and Gamma approximation to the null distribution.
For GC-VAR, we use a history of length $l=32$.
For TE-KSG, we set $l=1$---higher values required significantly more running time.
For a fair comparison, we parameterize all models such that events have impacts at lag 1.

\paragraph{Implementation.}
We implemented EITEST in Python, using the KS two-sample test from the
\texttt{SciPy} package\footnote{\url{http://www.scipy.org/}}, and the MMD two-sample
test provided by its authors\footnote{\url{http://www.gatsby.ucl.ac.uk/~gretton/mmd/mmd.htm}}.
For GC-VAR we used the implementation from the
\texttt{statsmodels}\footnote{\url{http://www.statsmodels.org/}} package.
TE-KSG was estimated with the \texttt{Java Information Dynamics Toolkit}
(JIDT)\footnote{\url{http://jlizier.github.io/jidt/}}.
Supplementary material and code can be found on \url{https://github.com/diozaka/eitest}.

\subsection{Simulation study.}

We now describe the three event impact models used for evaluation
and report the performances of all tests.
In the first model, events have impact on the mean of the time series, in the
second they modulate its variance, while in the third they alter the tails
of its distribution.
In all experiments, we first generate an event series of length $T$ with $N$ event
occurrences by sampling (without replacement) $N$ time steps $t_1$, ..., $t_N$ and setting
$E_{t_n} = 1$ for these time steps.

\paragraph{Impacts in mean.}
We modulate the mean of the time series by a moving average model \cite{Hamilton1994}
of order $q \in \mathbb{N}$ that uses events as innovations:
\begin{equation}
\label{eqnMeanImpacts}
X_t = \sum_{j=1}^{q} \phi_j E_{t-j} + Z_t.
\end{equation}
The weights $\boldsymbol{\phi} = [\phi_1, ..., \phi_q] \in \mathbb{R}^{q}$ determine
the shape of the event impacts and $Z_t \iid \mathcal{N}(0,1)$ is an error term.
We control the signal to noise ratio $r_\text{m}$ between event impacts and error term by sampling
$\boldsymbol{\phi} \sim \mathcal{N}(\boldsymbol{0}, r_\text{m} \cdot \boldsymbol{I}_{q})$.
In this model, every event has the same deterministic impact on the time series and
overlapping impacts simply add up.
Large values of $q$ introduce long-range temporal impacts that may lead to severe overlaps
and complicate the detection problem.

\paragraph{Impacts in variance.}
We modulate the variance of the time series by sampling from a normal distribution
with variance depending on the event series lagged by $q \in \mathbb{N}$ time steps:
\begin{equation}
\label{eqnVarImpacts}
X_t \mid E_{t-q} \sim \mathcal{N}(0, 1+r_\text{v} \cdot E_{t-q}).
\end{equation}
The factor $r_\text{v} > 0$ specifies the increase in variance induced by event occurrences.
The larger the value of $r_\text{v}$, the stronger the impacts, and the easier---at least
theoretically---the detection. By construction, the event impact model from
Equation~\ref{eqnVarImpacts} alters \emph{only} the variance of the distribution,
and no other property. In particular, the mean remains unchanged.

\paragraph{Impacts in tails.}
At last, we modulate the tail behavior of the time series by sampling either from
a normal distribution (light tails) or from Student's t-distribution (heavy tails),
depending on whether there was an event occurrence at lag $q$:
\begin{equation}
X_t \mid E_{t-q} \sim \begin{cases}
  \mathcal{N}\left(0, \frac{r_\text{t}}{r_\text{t}-2}\right), & \text{if } E_{t-q}=0,\\
  \operatorname{Student-t}(r_\text{t}), & \text{if } E_{t-q}=1
\end{cases}
\end{equation}
The parameter $r_\text{t} \ge 3$ specifies the degrees of freedom for Student's t-distribution.
A random variable $Z \sim \operatorname{Student-t}(r_\text{t})$ with $r_\text{t} \ge 3$
has mean $\expectation[Z] = 0$ and variance $\variance[Z] = \frac{r_\text{t}}{r_\text{t}-2}$.
Therefore, the model for impacts in tail behavior does not alter the mean or variance
of the time series. For $r_\text{t} \gg 3$, Student's t-distribution approximates
a normal distribution. Detection of event impacts is thus easiest for small
values of $r_\text{t}$ and becomes more difficult for larger values.

\paragraph{Benchmark and results.}
Our default parameterization for the event series is $T=8192$, with $N=128$ events in case
of the mean and variance impact models, and $N=1024$ for the tail impact model.
For the mean impact model we choose a default impact length of $q=8$ and signal-to-noise ratio
$r_\text{m}=10$. For the variance impact model we fix the delay at $q=1$ and set the default
variance increase to $r_\text{v}=4$. For the tail impact model we also fix the delay at $q=1$
and set the default degrees of freedom to $r_\text{t}=3$.
We change the detection difficulty by varying all parameters from these default
values. For every parameterization, we generate 100 pairs of coupled event series and
time series and 100 uncoupled pairs.

\begin{figure}[tbp]
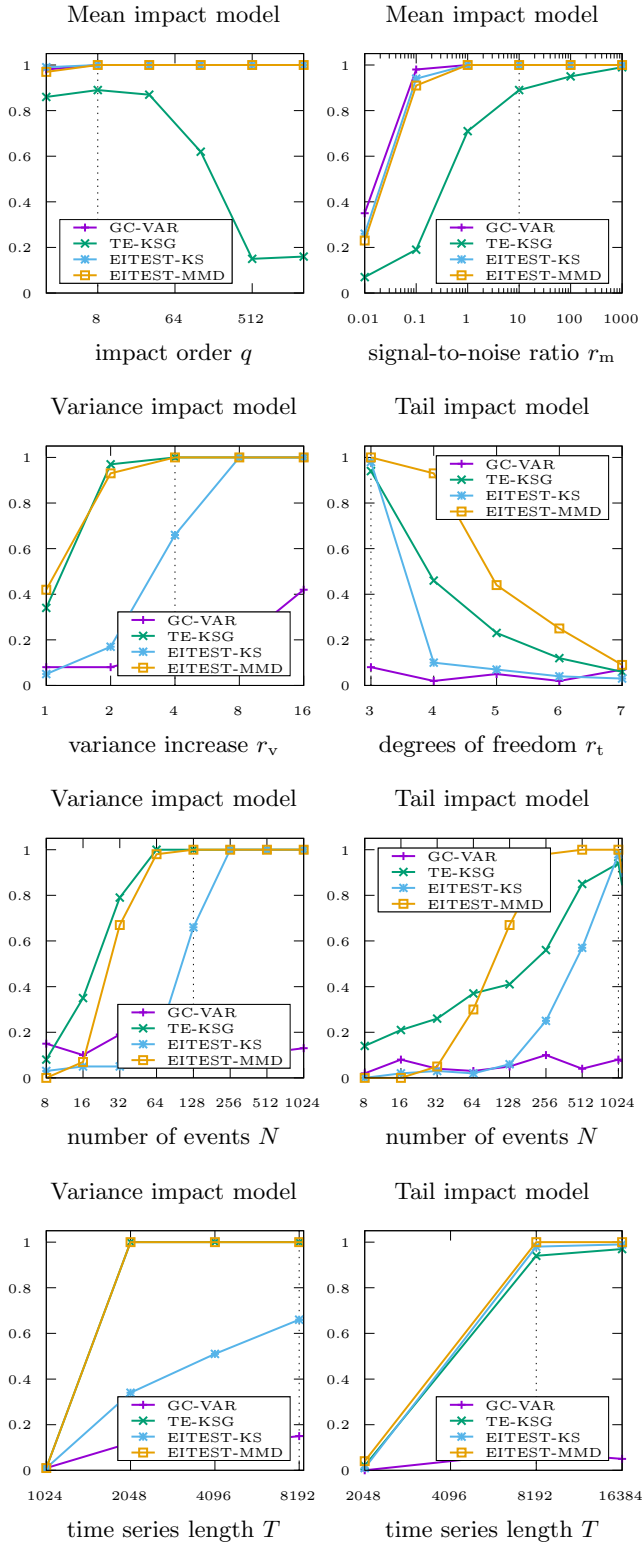

\begin{center}
\begin{gnuplot}[terminal=cairolatex,terminaloptions={size 3.33537,8.225}]
set multiplot layout 4,2
set format y '\tiny{set format x '\tiny{

set lmargin 3
set rmargin 1.5

#set ylabel '\tiny{power}'
set yrange [0:1.05]
set pointsize 0.5

#set tmargin at screen 0.9
#set bmargin at screen 0.625
set title '\small{Mean impact model}'
set key box opaque inside bottom left spacing .65 reverse Left samplen 1 width -2
set xlabel '\small{impact order $q$}'
set logscale x 2
set xrange [2:2048]
set xtics 1, 8
set arrow from 8, 0 to 8, 1.05 nohead dashtype 3
plot "data/benchmark-eigrte-MaVarTail.dat" i 0 u 1:4 w lp lw 3 t '\tiny{GC-VAR}', \
     "data/benchmark-eigrte-MaVarTail.dat" i 0 u 1:5 w lp lw 3 t '\tiny{TE-KSG}', \
     "data/benchmark-eigrte-MaVarTail.dat" i 0 u 1:2 w lp lw 3 t '\tiny{EITEST-KS}', \
     "data/benchmark-eigrte-MaVarTail.dat" i 0 u 1:3 w lp lw 3 t '\tiny{EITEST-MMD}'
unset arrow
unset title

set title '\small{Mean impact model}'
set key box opaque inside bottom right spacing .65 reverse Left samplen 1 width -2
set xlabel '\small{signal-to-noise ratio $r_\text{m}$}'
set logscale x 10
set xrange [0.01:1000]
set xtics 0.01, 10
set arrow from 10, 0 to 10, 1.05 nohead dashtype 3
plot "data/benchmark-eigrte-MaVarTail.dat" i 1 u 1:4 w lp lw 3 t '\tiny{GC-VAR}', \
     "data/benchmark-eigrte-MaVarTail.dat" i 1 u 1:5 w lp lw 3 t '\tiny{TE-KSG}', \
     "data/benchmark-eigrte-MaVarTail.dat" i 1 u 1:2 w lp lw 3 t '\tiny{EITEST-KS}', \
     "data/benchmark-eigrte-MaVarTail.dat" i 1 u 1:3 w lp lw 3 t '\tiny{EITEST-MMD}'
unset arrow
unset title

set title '\small{Variance impact model}'
set key box opaque inside bottom right spacing .65 reverse Left samplen 1 width -2
set xlabel '\small{variance increase $r_\text{v}$}'
set logscale x 2
set xrange [1:16]
set xtics 1, 2
set arrow from 4, 0 to 4, 1.05 nohead dashtype 3
plot "data/benchmark-eigrte-MaVarTail.dat" i 4 u 1:4 w lp lw 3 t '\tiny{GC-VAR}', \
     "data/benchmark-eigrte-MaVarTail.dat" i 4 u 1:5 w lp lw 3 t '\tiny{TE-KSG}', \
     "data/benchmark-eigrte-MaVarTail.dat" i 4 u 1:2 w lp lw 3 t '\tiny{EITEST-KS}', \
     "data/benchmark-eigrte-MaVarTail.dat" i 4 u 1:3 w lp lw 3 t '\tiny{EITEST-MMD}'
unset arrow
unset title

set title '\small{Tail impact model}'
set key box opaque inside top right spacing .65 reverse Left samplen 1 width -2
set xlabel '\small{degrees of freedom $r_\text{t}$}'
unset logscale x
set xrange [2.9:7]
set xtics 3, 1
set arrow from 3, 0 to 3, 1.05 nohead dashtype 3
plot "data/benchmark-eigrte-MaVarTail.dat" i 7 u 1:4 w lp lw 3 t '\tiny{GC-VAR}', \
     "data/benchmark-eigrte-MaVarTail.dat" i 7 u 1:5 w lp lw 3 t '\tiny{TE-KSG}', \
     "data/benchmark-eigrte-MaVarTail.dat" i 7 u 1:2 w lp lw 3 t '\tiny{EITEST-KS}', \
     "data/benchmark-eigrte-MaVarTail.dat" i 7 u 1:3 w lp lw 3 t '\tiny{EITEST-MMD}'
unset arrow
unset title

set title '\small{Variance impact model}'
set key box opaque inside bottom right spacing .65 reverse Left samplen 1 width -2
set xlabel '\small{number of events $N$}'
set logscale x 2
set xrange [8:1024]
set xtics 8, 2
set arrow from 128, 0 to 128, 1.05 nohead dashtype 3
plot "data/benchmark-eigrte-MaVarTail.dat" i 2 u 1:4 w lp lw 3 t '\tiny{GC-VAR}', \
     "data/benchmark-eigrte-MaVarTail.dat" i 2 u 1:5 w lp lw 3 t '\tiny{TE-KSG}', \
     "data/benchmark-eigrte-MaVarTail.dat" i 2 u 1:2 w lp lw 3 t '\tiny{EITEST-KS}', \
     "data/benchmark-eigrte-MaVarTail.dat" i 2 u 1:3 w lp lw 3 t '\tiny{EITEST-MMD}'
unset arrow
unset title

set title '\small{Tail impact model}'
set key box opaque inside top left spacing .65 reverse Left samplen 1 width -2
set xlabel '\small{number of events $N$}'
set logscale x 2
set xrange [8:1100]
set xtics 8, 2
set arrow from 1024, 0 to 1024, 1.05 nohead dashtype 3
plot "data/benchmark-eigrte-MaVarTail.dat" i 5 u 1:4 w lp lw 3 t '\tiny{GC-VAR}', \
     "data/benchmark-eigrte-MaVarTail.dat" i 5 u 1:5 w lp lw 3 t '\tiny{TE-KSG}', \
     "data/benchmark-eigrte-MaVarTail.dat" i 5 u 1:2 w lp lw 3 t '\tiny{EITEST-KS}', \
     "data/benchmark-eigrte-MaVarTail.dat" i 5 u 1:3 w lp lw 3 t '\tiny{EITEST-MMD}'
unset arrow
unset title

set title '\small{Variance impact model}'
set key box opaque inside bottom right spacing .65 reverse Left samplen 1 width -2
set xlabel '\small{time series length $T$}'
set logscale x 2
set xrange [1024:8500]
set xtics 256, 2
set arrow from 8192, 0 to 8192, 1.05 nohead dashtype 3
plot "data/benchmark-eigrte-MaVarTail.dat" i 3 u 1:4 w lp lw 3 t '\tiny{GC-VAR}', \
     "data/benchmark-eigrte-MaVarTail.dat" i 3 u 1:5 w lp lw 3 t '\tiny{TE-KSG}', \
     "data/benchmark-eigrte-MaVarTail.dat" i 3 u 1:2 w lp lw 3 t '\tiny{EITEST-KS}', \
     "data/benchmark-eigrte-MaVarTail.dat" i 3 u 1:3 w lp lw 3 t '\tiny{EITEST-MMD}'
unset arrow
unset title

set title '\small{Tail impact model}'
set key box opaque inside bottom right spacing .65 reverse Left samplen 1 width -2
set xlabel '\small{time series length $T$}'
set logscale x 2
set xrange [2048:16384]
set xtics 2048, 2
set arrow from 8192, 0 to 8192, 1.05 nohead dashtype 3
plot "data/benchmark-eigrte-MaVarTail.dat" i 6 u 1:4 w lp lw 3 t '\tiny{GC-VAR}', \
     "data/benchmark-eigrte-MaVarTail.dat" i 6 u 1:5 w lp lw 3 t '\tiny{TE-KSG}', \
     "data/benchmark-eigrte-MaVarTail.dat" i 6 u 1:2 w lp lw 3 t '\tiny{EITEST-KS}', \
     "data/benchmark-eigrte-MaVarTail.dat" i 6 u 1:3 w lp lw 3 t '\tiny{EITEST-MMD}'
unset arrow
unset title
\end{gnuplot}
\caption{True positive rates of EITEST, GC-VAR and TE-KSG for the mean, variance
and tail impact models.}
\label{figPerformance}
\end{center}
\end{figure}

Figure~\ref{figPerformance} shows the true positive rates of all competing tests.
EITEST outperforms or is on par with all approaches almost across the whole model parameter space.
EITEST-MMD generally outperforms EITEST-KS, possibly due to a
higher statistical power of the MMD two-sample test compared to the KS two-sample test
for small sample sizes.
Despite being nonparametric, EITEST-MMD is on par with the parametric GC-VAR test on
\emph{impacts in mean}. TE-KSG, which is also nonparametric, fails to detect higher order impacts
in mean.
As expected, GC-VAR does not detect \emph{impacts in variance or tails}, whereas EITEST-MMD and
TE-KSG are sensitive in these two scenarios as well. In the case of tail impacts, EITEST-MMD
outperforms TE-KSG and GC-VAR by a large margin.
TE-KSG appears more powerful than EITEST-MMD for impacts in tail and variance when the number
of events is small.
This effect may be explained by the short history length $l=1$ for TE-KSG
(compared to $K=32$ for EITEST), which makes estimation of transfer entropy easier.
However, for $N\ge 64$ events, EITEST-MMD reaches and surpasses the performance of TE-KSG.
In summary, EITEST-MMD is the only approach that reliably detects all three types of impacts.
As a sanity check, we provide the false positive rates of the tests in the online supplementary material.
We observe that in our simulation study all tests approximately control the false positive rate
at the desired significance level $\alpha=.05$.
There is a slight tendency of EITEST-MMD to over-reject (false positive rates above the controlled
level $\alpha$). Since we do not observe this behavior in EITEST-KS, we suspect this
behavior is due to the Gamma approximation to the MMD null distribution.

\subsection{Application: Electricity monitoring.}
\label{secAMPds}
We now use our test for household electricity monitoring in a smart home environment.
Specifically, we analyze the effect of turning on the clothes washer
on various electricity meters in a residential house.

\begin{figure}[tb]
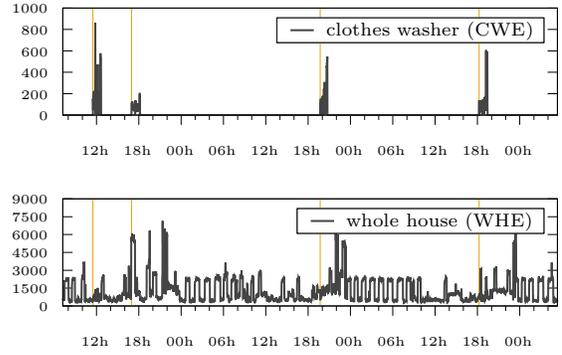

\begin{center}
\begin{gnuplot}[terminal=cairolatex,terminaloptions={size 3.33537,2.0}]
set multiplot layout 2,1
set format y '\tiny{set format x '\tiny{set timefmt "set xdata time
set xrange [1333523600:1333776000]
set key box opaque inside top right reverse Left samplen 1 width -5
set xtics out nomirror
plot "data/ampds-cwe-ev.csv" u ($1-25200):($2*1000) w i lc 4 lw 1 notitle, \
     "data/ampds-all.csv" u ($1-25200):8 w l lc rgb '#444444' lw 3 title '\scriptsize{clothes washer (CWE)}'
set ytics 0, 1500
set yrange [0:9000]

set key box opaque inside top right reverse Left samplen 1 width -3.5
plot "data/ampds-cwe-ev.csv" u ($1-25200):($2*10000) w i lc 4 lw 1 notitle, \
     "data/ampds-all.csv" u ($1-25200):2 w l lc rgb '#444444' lw 3 title '\scriptsize{whole house (WHE)}'
\end{gnuplot}
\caption{Clothes washer and whole house electricity consumption with clothes washing
events (orange).}
\label{figCWE}
\end{center}
\end{figure}

\paragraph{Data.} For the experiment, we use the publicly available
Almanac of Minutely Power dataset (AMPds)~\cite{Makonin2016}.
The dataset contains two years of minutely electricity, water and natural gas measurements
from a residential house in Canada.
We focus on electricity consumption, which was recorded using 21 physical meters placed
at various locations in the building to separately measure the consumption of different
household appliances (clothes washer, clothes dryer, dishwasher, etc.), rooms (bedroom,
home office, garage, etc.), and the whole house consumption.
Each time series contains 1,051,200 measurements.
We extract 413 clothes washing events from the clothes washer electricity (CWE) meter.
An excerpt of the resulting event series is depicted in Figure~\ref{figCWE} along with
the clothes washer meter (CWE, left) and the whole house meter (WHE, right)
between April 4th, 2012 and April 7th, 2012.
The different scales of the y-axes indicate the low signal to noise ratio of the clothes
washer impacts within the whole house time series, which makes the detection problem hard.

\paragraph{Results.}
In all experiments, we set the maximum lag to $K=120$ minutes (2 hours).
The p-values obtained on all meters are shown in Table~\ref{tblAMPds}.
Results that are significant at level $\alpha=.05$ (unadjusted) are shaded.
Since the time series are very long, neither GC-VAR nor TE-KSG terminated within one hour and had
to be aborted.
The MMD-based test rejects on all instances where the KS-based tests rejects, and some more.
This behavior confirms that EITEST-MMD is more powerful than EITEST-KS.
Despite the low signal to noise ratio, EITEST-KS and EITEST-MMD correctly identify a statistically
significant association between the clothes washer and the whole house meter (WHE).
Furthermore, the tests identify statistically significant associations in several other meters, e.g.,
the clothes dryer meter (CDE).
All of these meters can potentially be used to detect clothes washing events.
Since the time series are univariate, we can visualize the post-event behavior
$F^k_{1,0,...,0}$ for all meters at increasing lags $k$ to get insights into
the nature of these associations and build a suitable event detection algorithm.
Visualizations can be found in the online supplementary material.

\begin{table}[tb]
\centering
\caption{AMPds p-values}
\label{tblAMPds}
\scriptsize
\begin{tabular}{c r r r r}
\toprule
meter & EITEST-KS & EITEST-MMD & GC-VAR & TE-KSG \\
\midrule
WHE &  \sigcell{${}<.0001$} &\sigcell{${}<.0001$} & \multicolumn{2}{c}{\multirow{23}{*}{no results}} \\
RSE &           $   .9999$  &         $   .9721$  & &   \\
GRE &           $   .9999$  &         $   .8754$  & &   \\
MHE &  \sigcell{${}<.0001$} &\sigcell{${}<.0001$} & &   \\
B1E &           $   .9999$  &         $   .9819$  & &   \\
BME &           $   .8629$  &\sigcell{${}<.0001$} & &   \\
CWE &  \sigcell{${}<.0001$} &\sigcell{${}<.0001$} & &   \\
DWE &           $   .9999$  &         $   .9759$  & &   \\
EQE &           $   .9999$  &\sigcell{$   .0119$} & &   \\
FRE &           $   .9999$  &         $   .9998$  & &   \\
HPE &           $   .9999$  &\sigcell{$   .0152$} & &   \\
OFE &           $   .9999$  &         $   .6240$  & &   \\
UTE &           $   .9999$  &\sigcell{$   .0074$} & &   \\
WOE &           $   .9999$  &         $   .9340$  & &   \\
B2E &  \sigcell{$   .0045$} &\sigcell{${}<.0001$} & &   \\
CDE &  \sigcell{${}<.0001$} &\sigcell{${}<.0001$} & &   \\
DNE &           $   .9999$  &         $   .9728$  & &   \\
EBE &           $   .9999$  &         $   .0562$  & &   \\
FGE &           $   .9999$  &         $   .9313$  & &   \\
HTE &  \sigcell{${}<.0001$} &\sigcell{${}<.0001$} & &   \\
OUE &  \sigcell{${}<.0001$} &\sigcell{${}<.0001$} & &   \\
TVE &           $   .9999$  &         $   .3944$  & &   \\
UNE &  \sigcell{$   .0084$} &\sigcell{$   .0004$} & &   \\
\bottomrule
\end{tabular}
\end{table}

\subsection{Application: Earthquakes on Twitter.}
\label{secCrimson}

At last, we analyze the coupling between earthquakes and German social media usage.
Since social media reactions often come in bursts of posts, we expect that events temporarily
fatten the tails of the conditional distributions.
We first test whether daily usage of the keyword ``earthquake'' in German Twitter is influenced by
the occurrence of severe earthquakes worldwide. We then focus specifically on earthquakes that hit
China, the country with the largest number of disastrous earthquakes in the time period we study.

\begin{figure}[tb]
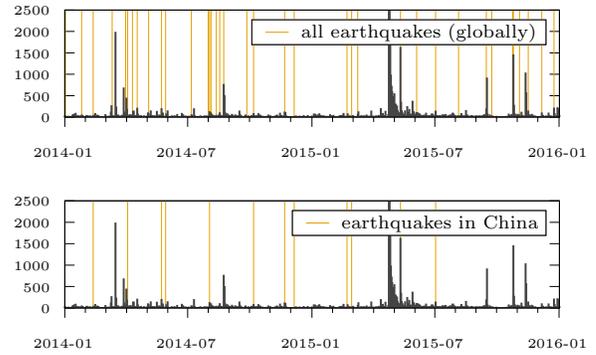

\begin{center}
\begin{gnuplot}[terminal=cairolatex,terminaloptions={size 3.33537,2.0}]
set multiplot layout 2,1
set format y '\tiny{set format x '\tiny{set timefmt "set xdata time
set xrange ["2014-01-01":"2016-01-02"]
set xtics "2010-01-01", 6*2629746, "2016-12-31" out nomirror
set yrange [0:2500]
set key box opaque inside top right reverse Left samplen 1 width -9
#set title '\scriptsize{German keyword usage and global earthquakes}'
plot "data/emdat-earthquake-all.csv" u 1:($2*5000) w i lc 4 lw 1 title '\scriptsize{all earthquakes (globally)}', \
     "data/twitter-earthquake-germany.csv" u 1:2 w i lc rgb '#444444' lw 3 notitle

set key box opaque inside top right reverse Left samplen 1 width -6
#set title '\scriptsize{German keyword usage and earthquakes in China}'
plot "data/emdat-earthquake-china.csv" u 1:($2*5000) w i lc 4 lw 1 title '\scriptsize{earthquakes in China}', \
     "data/twitter-earthquake-germany.csv" u 1:2 w i lc rgb '#444444' lw 3 notitle
\end{gnuplot}
\caption{Volume of the keyword ``earthquake'' in German Twitter over time, along with two earthquake event series.}
\label{figEarthquakes}
\end{center}
\end{figure}

\paragraph{Data.}
We obtained time series of the daily number of tweets posted in Germany that contain the keyword
``earthquake'', translated into more than 30 languages, between 2010 and 2017 (2,557 days),
using Crimson Hexagon's ForSight platform.\footnote{\url{https://www.crimsonhexagon.com/}}
For the daily earthquake event series, we used the publicly available Emergency Events Database
(EM-DAT) provided by the Centre for Research on the Epidemiology of Disasters (CRED)\footnote{\url{http://emdat.be/}}
and extracted all severe earthquakes in the same time period. We created two event series: the first
containing all earthquakes globally (162 events), the second containing only earthquakes
in China (40 events).
Excerpts from the two pairs are depicted in Figure~\ref{figEarthquakes}.

\paragraph{Results.}
We set the maximum lag to $K=7$ days.
According to EITEST-KS and EITEST-MMD, the event series with all global earthquakes is coupled with
German Twitter activity: for both variants, the null hypothesis of independence is rejected with $p < .0001$.
This result matches the intuition that there should be an association between the series.
GC-VAR does not detect an association ($p=.1919$).
When it comes to the event series with earthquakes in China, our tests do not find enough evidence for
a statistical association (EITEST-KS: $p=.4090$, EITEST-MMD: $p=.4225$), which may indicate a lack of
awareness of these events in the German public. However, GC-VAR detects an association ($p=.0090$)
and thus contradicts its earlier result.
TE-KSG provides inconsistent results on both tasks: the test delivers largely fluctuating p-values when run repeatedly. Overall, the results on earthquakes in China are inconclusive.
A visualization of the post-event behavior of the time series for both event series can
be found in the online supplementary material.

\section{Conclusions}

Our event information test (EITEST) is designed to test for shared information
between a time series and an event series in a nonparametric way. The ultimate
goal is to identify time series that can be exploited for event detection.
We reduce the independence testing problem to a problem of multiple two-sample testing.
This reduction allows us to apply recent approaches to nonparametric two-sample testing.
In particular, with EITEST-MMD, associations can be assessed for time series of arbitrary
domains, as long as a suitable kernel for the MMD statistic is available.
Since EITEST itself has only a single intuitive parameter, it is easy to apply in practice.
Our simulations show that EITEST outperforms or is on par with methods for causal
inference in detecting relevant statistical associations,
and is the only approach that reliably detects all three kinds of event impact
that we tested for. As it is linear in the time series length $T$, it can be applied
to very long input sequences, where existing tests fail to deliver results within
a reasonable amount of time.

\balance

\bibliographystyle{abbrv}
\bibliography{library}

\end{document}